\documentclass[twocolumn,showpacs,preprintnumbers,amsmath,amssymb]{revtex4}
\usepackage{tabularx,graphicx}

\usepackage{color}
\usepackage{hyperref}
\hypersetup{
    colorlinks=true,
    linkcolor=blue,
    filecolor=blue,      
    urlcolor=blue,
}

\begin{document}

\newcommand{\beq}{\begin{equation}}
\newcommand{\eeq}{\end{equation}}
\newcommand{\beqn}{\begin{eqnarray}}
\newcommand{\eeqn}{\end{eqnarray}}
\newcommand{\bmath}{\begin{subequations}}
\newcommand{\emath}{\end{subequations}}
\newcommand{\bra}[1]{\langle #1|}
\newcommand{\ket}[1]{|#1\rangle}

\title{Towards an understanding of  hole superconductivity}
\author{J. E. Hirsch }
\address{Department of Physics, University of California, San Diego,
La Jolla, CA 92093-0319}

\begin{abstract} 
From the very beginning K. Alex M\"uller emphasized that the materials he and George Bednorz discovered
in 1986  were $hole$ superconductors.  
Here I would like to share with him
and others  what I believe to be $the$ key reason for why high $T_c$ cuprates as well as all other superconductors are hole superconductors, which I only came to understand a few months ago. 
This paper is dedicated to Alex  M\"uller on the occasion of his 90th birthday.
 \end{abstract}
\pacs{}
\maketitle
 \section{introduction}
The very first paper by K. Alex M\"uller listed  in Web of Science, from   when he was a 
youthful 27-year-old, is on an apparatus to measure Hall effect\ \cite{alex1}. It shows that from the very beginning of his scientific career, Alex M\"uller was
well  aware of the difference between electrons and holes. \cite{wiki}

          \begin{figure}
 \resizebox{8.5cm}{!}{\includegraphics[width=6cm]{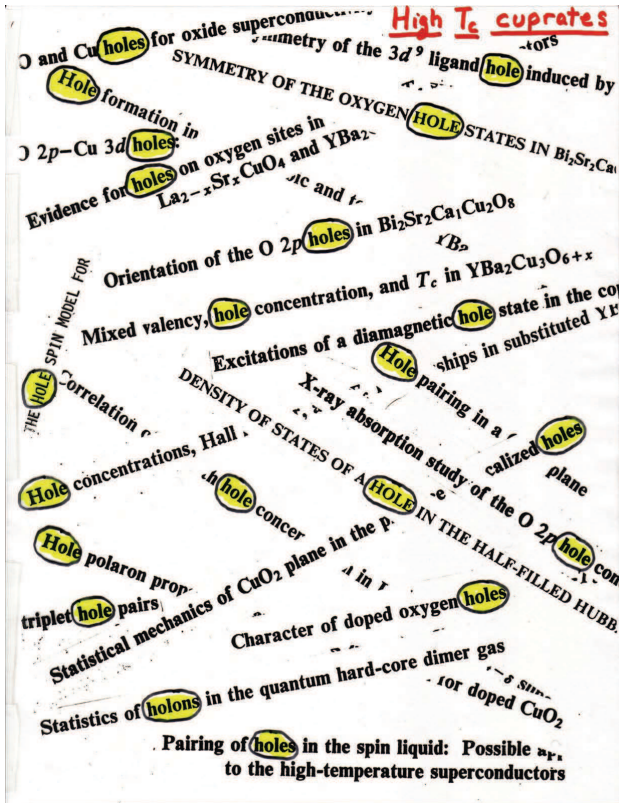}}
 \caption { Titles of some papers published in the early days of high $T_c$ research (see text). }
 \label{figure1}
 \end{figure}

The concept of holes has played a prominent role  in semiconductor physics for a long time, as exemplified by the title of Shockley's 1950 book
``Electrons and holes in semiconductors''. It also of course has played a prominent role in quantum electrodynamics since Dirac discovered holes in 1930. But it had played 
essentially no 
role in superconductivity until the discovery of high $T_c$ cuprates. 

In his 1987 paper in LT18, `A Road towards High $T_c$  Superconductivity' \cite{alex2} Alex spelled it out clearly for the first time: {\it ``Basically, all these materials, are hole superconductors.''}
Again in his 1987 Science paper
`The Discovery of a Class of High-Temperature Superconductors' \cite{alex3} Alex remarks 
 {\it   ``This new class of materials found at the IBM Zurich Research Laboratory are hole rather than electron superconductors''}. 
 In his 1988 paper in the proceedings of the NEC Symposium on Mechanisms of High Temperature Superconductivity he states \cite{alexnec}
 {\it ``As the $T_c$ of hole-containing $BaBiO_3$ is more than twice 
as high as that of the electron-containing compound, 
one might expect an enhancement of $T_c$ for 
\textbf {hole superconductivity} over electron superconductivity 
in the cuprates if the latter are found.''} It is clear that Alex was struck   by the realization that holes seemed to be favorable for
superconductivity, which was not part of his initial  theoretical views
on high temperature superconductivity (Jahn-Teller polarons)  that guided him and Bednorz in their  search and led them to their discovery. Thus, he   
emphasized the `hole' aspect   in 
many of his early papers  and talks. The  era of hole superconductivity had begun.

Soon thereafter, everybody working on cuprates was talking about `holes'. Figure 1 shows a transparency I made in those early days with random titles of papers
that I used in talks to emphasize this point.

I met Alex M\"uller for the first time at the NEC symposium in late 1988 and listened keenly to his talk, one of the first in the
program. I still have the handwritten notes I took at the time, 6 pages of them. In the middle of the second page there is the
statement ``These materials are hole superconductors'', with ``hole superconductors'' underlined. I still remember vividly the emphasis
he put on those words in his presentation, that  deeply impressed me at the time.

 Later in that meeting, H. Takagi made a comprehensive presentation of transport properties of $(La_{1-x}Sr)_2CuO_4$ \cite{takagi}. 
He showed  a slide of $T_c$ versus hole concentration ($p$)  and asked 
``why does superconductivity disappear'' at  $p=0.15$?''
On the very next slide he showed a graph of Hall coefficient versus hole concentration showing that it 
changes sign from positive to negative precisely at $p=0.15$.

Takagi suggested  in this presentation   that $T_c$ goes to zero in the overdoped regime because of a cross-over from
Mott-Hubbard to Fermi liquid regime. He did $not$, to the best of my
recollection and according to my notes taken at the time, directly connect the change in sign of the
Hall coefficient from positive to negative to the disappearance of superconductivity.
I wondered for a long time why he hadn't done that and only much later I learned
why \cite{source}. It turns out that at that time, October 1988, he and his coworkers already
had discovered the so-called `electron superconductors'. He did not mention
this discovery at that meeting nor did other speakers, those results were announced
in January 1989 \cite{edoped}. But this clearly must have been the reason why he did not think  that the type of charge carrier
(whether hole or electron) was a determining factor.
It took many more years and a lot of experiments to establish that the
electron-doped cuprates are in fact also {\it hole superconductors} \cite{greene}.

In the paragraph above I said the concept of holes had played $essentially$ no role in superconductivity. 
The caveat is because in fact  several researchers in the early days of superconductivity did suggest that 
a positive Hall  coefficient was favorable to superconductivity \cite{chapnik}. However, the concept fell completely
out of favor after the establishment of the BCS theory of superconductivity, within which the
character of the carriers, whether electrons or holes, plays no role.

 Ever since I heard that fateful talk by Alex M\"uller in 1988 I have been convinced that hole carriers are essential for superconductivity
 in all materials  \cite{holefirst}, not just in high $T_c$ cuprates. Together with Frank Marsiglio and other coworkers we have presented many arguments
 and calculations in favor of this hypothesis \cite{holesc}. 
 In this short paper I would like to  discuss what I think is  the most fundamental reason why holes are 
 indispensable  for superconductivity, that I have only understood  a few months ago. But first some preliminaries.

 \section{holes in condensed matter physics}
 The concept of holes in   solids was introduced by Werner Heisenberg in 1931 \cite{heisenberg}. He writes:
  {\it ``Die Elektrizitl\"atsleitung
in Metallen mit einer geringen Anzahl von
L\"ochern kann also in jeder Beziehung beschrieben werden wie
die Leitung in Metallen mit einer geringen Anzahl von positiven
Leitungselektronen. Daraus folgt unmittelbar der anomale
Halleffekt fur solche Metalle.''}
   Similarly Peierls in 1932 writes \cite{peierls}
  {\it ``Ein Band, in dem sich nur wenige Elektronen befinden, verhalten sich
in jeder Beziehung genau so, wie ein Band, in dem nur f\"ur wenige Elektronen
noch Platz ist, mit dem Unterschied, dass den freien Pl\"atzen
eine umgekehrte -- also positive -- Ladung zuzuschreiben ist. Da jedoch
die Leitf\"ahigkeit unabh\"angig vom Vorzeichen der Ladung ist, wird sich
dieser Unterschied zun\"achst in der Leitf\"ahigkeit noch nicht bernerkbar
machen.''} The reason for the `anomalous' (positive) Hall coefficient in  metals 
with `hole' carriers was worked out by Peierls upon
the suggestion of Heisenberg already in 1929 \cite{peierlshall}.
  
  Both of the above statements say that {\it ``in jeder Beziehung''}, i.e. {\it ``in every respect''}, holes are just like electrons in solids.
  This point of view  has been pervasive  in condensed matter physics ever since. 
  Yet it is incorrect. If it was correct, there would be no superconductivity.
  
  In a paper I wrote in 2005 I listed many reasons why holes are $not$ like electrons, as shown in Figure 2. 
  The most relevant one regarding superconductivity is highlighted. I will explain this in a later section.
  
    \begin{figure} [htb]
 \resizebox{8.5cm}{!}{\includegraphics[width=6cm]{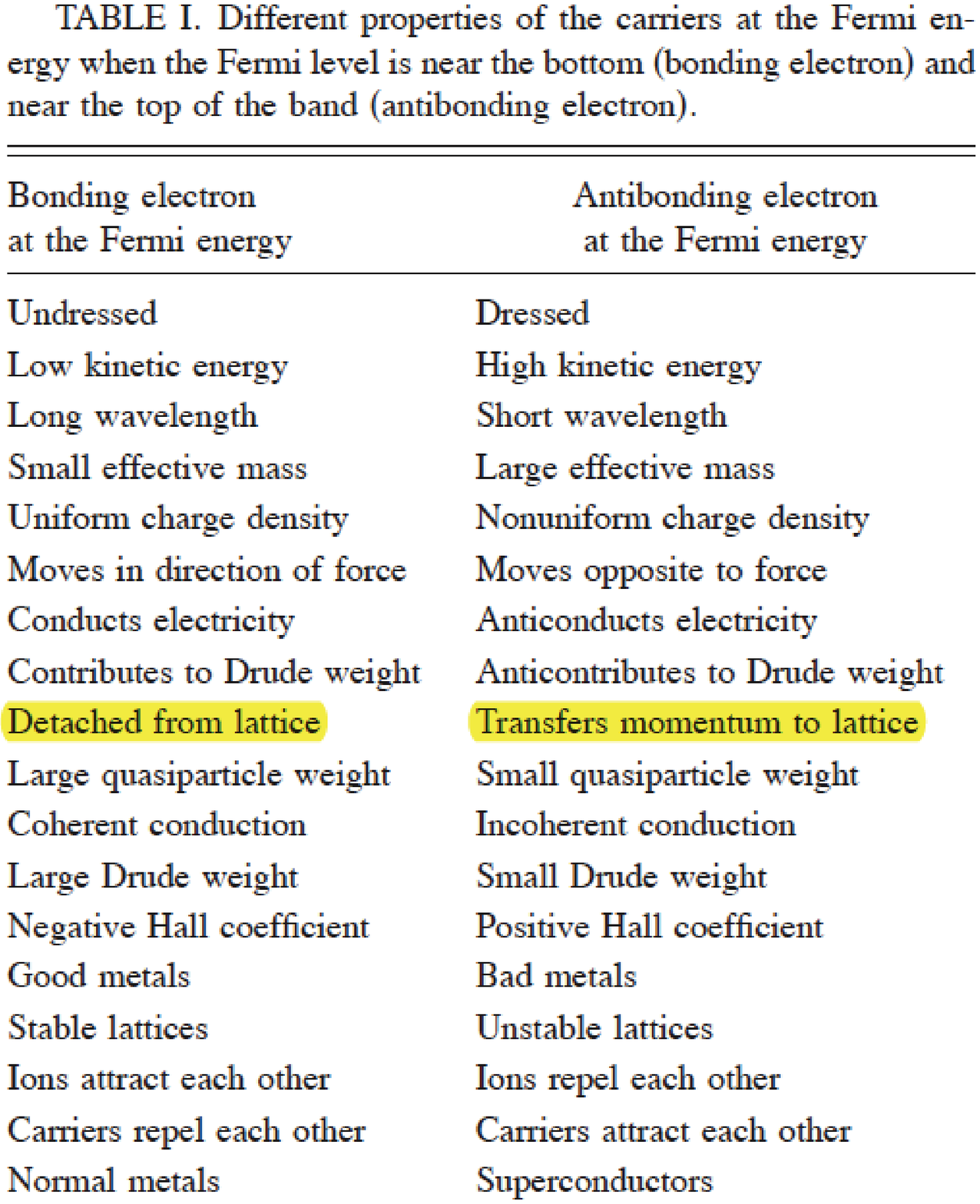}}
 \caption { From a paper the author wrote in 2005 \cite{holeelec2}. The most important reason for why
 holes are necessary for superconductivity is highlighted. }
 \label{figure1}
 \end{figure}

 \section{ electron-hole asymmetric polarons and dynamic Hubbard models }

Alex M\"uller has always focused on polarons as being at the root of high temperature superconductivity \cite{alexfirst,alexpolarons,alexpolarons2}.
Somewhat overlapping with his idea, within the theory of hole superconductivity, high $T_c$ originates
in small {\it electron-hole asymmetric} polarons \cite{tang,ehpol}. In our view the dominant polaronic aspect arises from electron-electron interactions
rather than from electron-lattice interactions. Nevertheless, even for electron-phonon polarons electron-hole asymmetry can play a big role
and favor hole over electron superconductivity \cite{holepolarons}, consistent with Alex's expectations.

A very simple, natural and general extension of the conventional Hubbard model leads to `dynamic Hubbard models' and 
electron-hole asymmetric electronic polarons. In the conventional Hubbard model, electrons in doubly
occupied orbitals pay the Coulomb repulsion price $U$, but their orbitals are unmodified relative to the singly-occupied orbital. However,
in reality  a doubly occupied atomic orbital expands relative to the singly-occupied orbital,
due to electron-electron repulsion. This `orbital relaxation' causes a reduction of the bare $U$ and leads to electron-hole asymmetry.
Dynamic Hubbard models describe this physics  \cite{dynh}.  When a hole propagates, the orbital relaxation causes `dressing' of the quasiparticle and
effective mass enhancement, as in small polarons. Instead, when an electron propagates no such effects exist. The effects are largest when
the effective ionic charge is small, so that the modification of the orbital upon double occupancy is large. Figure 3 shows a `cartoon picture'
of the physics  that is included in dynamic Hubbard models and not in the conventional Hubbard  model. The orbital expansion lowers the electronic kinetic energy (as well as the Coulomb repulsion), and causes negative charge to expand outward. Both aspects are relevant to the physics of hole
superconductivity \cite{holescripta}.

    \begin{figure} [htb]
 \resizebox{8.5cm}{!}{\includegraphics[width=6cm]{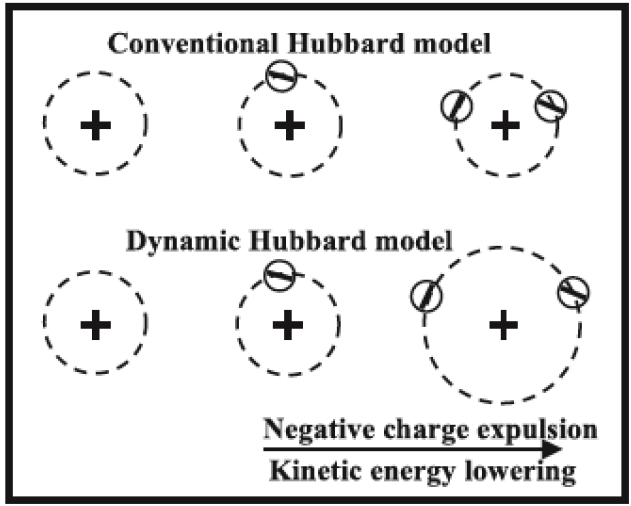}}
 \caption { Schematics of the physics described by dynamic Hubbard models}
 \label{figure1}
 \end{figure} 
 
    \begin{figure} [htb]
 \resizebox{8.5cm}{!}{\includegraphics[width=6cm]{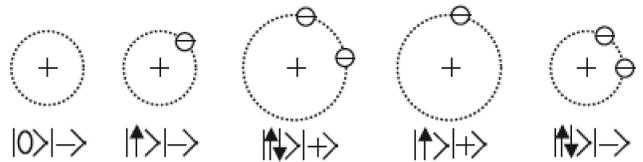}}
 \caption { Site states of dynamic Hubbard model with an auxiliary spin degree of freedom with states
 $|+>, |->$. The left three states (as well as $|\downarrow>|->$)  are  lowest in energy and are  the quasiparticle states in the low energy effective
 Hamiltonian with a correlated hopping term.}
 \label{figure1}
 \end{figure}

    \begin{figure} [htb]
 \resizebox{8.5cm}{!}{\includegraphics[width=6cm]{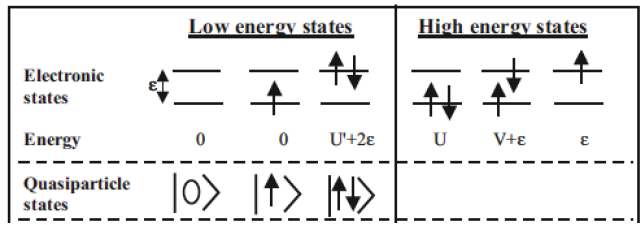}}
 \caption { Site states of dynamic Hubbard model with two orbitals per site. The low energy physics is identical
 to the one described by Fig. 4.}
 \label{figure1}
 \end{figure}

Various versions of dynamic Hubbard models can be constructed to embody this physics, involving auxiliary spin or local vibrational degrees of freedom,
or in a purely electronic version a tight binding model with two rather than one orbital per site \cite{dynhversions},
as shown in Figs. 4 and 5.  The physics of all these models is very similar. Figure 6 illustrates the fact
that holes have more difficulty propagating than electrons in these models. The low energy effective Hamiltonian that results from these models has a `correlated hopping' term $\Delta t$ 
that gives different hopping amplitudes depending on the occupation of the sites involved in the
hopping process and leads to kinetic energy driven pairing and
superconductivity when the Fermi level is close to the top of the band, i.e. for hole carriers \cite{holebcs}.
 The $T_c$ versus hole concentration dependence gives the bell-shaped behavior
 characteristic of the cuprates as well as the $T_c$ versus e/a (electron/atom) ratio in transition metal
 alloys \cite{tm} (Matthias' rules) \cite{matthias}.

    \begin{figure} [htb]
 \resizebox{8.5cm}{!}{\includegraphics[width=6cm]{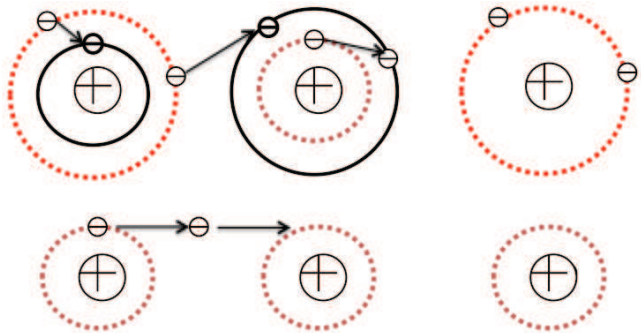}}
 \caption { Propagation of holes (upper pictures) versus propagation of electrons (lower pictures). 
 Holes are highly dressed because they cause a large disruption in their environment as they propagate, electrons are undressed. In the upper picture the ful (dotted) circles denote the orbital in the final (initial) state.}
 \label{figure1}
 \end{figure}

 The applicability of these models to high $T_c$ cuprates rests  on the assumption that doped holes
 go into oxygen $p\pi$ orbitals in the plane \cite{tang}, rather than $p\sigma$ orbitals as generally assumed.
 This assumption is supported by quantum chemical calculations by Goddard and coworkers \cite{goddard}.
 We have recently argued that band structure calculations get this wrong because they do not take into
 account the energy lowering that results from local orbital relaxation when a hole
 goes into the doubly occupied $O^{--}$ $p\pi$ orbital \cite{cuo}.
 These models also give rise to a strong tendency to charge inhomogeneity and phase
 separation due to the dominance of kinetic over potential energy \cite{cuo},
 which correlates with behavior found in the cuprates  \cite{disorder}.
 
 \section{hole superconductivity in materials}
 The models discussed in the previous section were introduced motivated by the physics of high $T_c$ cuprates.
We found that they describe in a very natural way several salient properties of cuprate superconductors
\cite{holebcs,marsiglio,holesc}, 
 in particular their:
  
  (i) Dome-like $T_c$ versus hole concentration dependence
  
  (ii) Positive pressure dependence of $T_c$
  
  (iii) Crossover between strong and weak coupling regimes as the hole concentration increases
  
  (iv) Crossover from incoherent to coherent behavior both as the hole concentration increases and as superconductivity
  sets in
  
  (v) Tunneling asymmetry, with larger current for negatively biased sample
  
  (vi) Apparent violation of conductivity sum rule, and transfer of optical spectral weight from high frequencies
  to low frequencies as superconductivity sets in.
  
  In addition, we have argued \cite{matmech} that these models lead to hole pairing and superconductivity in  the following classes \cite{classes}
   of superconducting materials:
  
  (1) Hole-doped cuprates  \cite{holebcs,marsiglio}
  
  (2) Electron-doped cuprates \cite{holesedoped}
  
  (3) Magnesium diboride   \cite{mgb2}
  
  (4) Transition metal series alloys \cite{tm}
  
  (5) Iron pnictides \cite{pnic}
  
  (6) Iron selenides  \cite{matmech}
  
  (7) Doped semiconductors  \cite{matmech,dopedsc}
  
  (8) Elements under high pressure \cite{hamlin,cava}
  
  (9) Sulphur hydride \cite{h2s}
  
  (10) A-15 materials \cite{a15,all}
  
  (11) All other superconductors \cite{all}
  
  For the simplest materials, the elements, there is an obvious preponderance of positive Hall coefficient 
  for superconducting elements and negative Hall coefficients for nonsuperconducting 
  elements \cite{chapnik,correlations}, as shown in Fig. 7.
 In the following we discuss the most fundamental reason that we believe makes holes 
 indispensable for superconductivity.

      \begin{figure} [htb]
 \resizebox{8.5cm}{!}{\includegraphics[width=6cm]{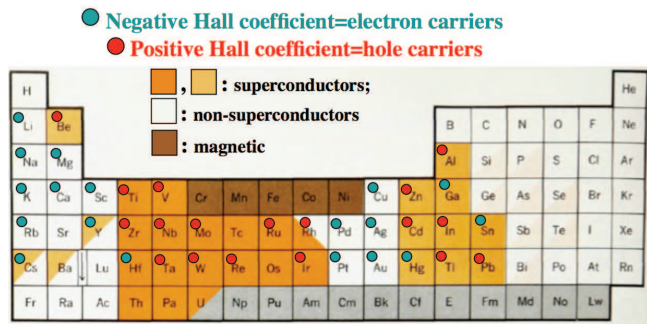}}
 \caption { Periodic table, showing the preponderance of superconductors among positive Hall
 coefficient elements and non-superconductors among negative Hall coefficient elements}
 \label{figure1}
 \end{figure} 

\section{the central question in superconductivity}
I would like to propose that the central question in the phenomenon of superconductivity is a very
basic one that even a child could ask, but scientists have  $never$  asked, nor answered. That question is the following:

{\it When a superconductor in a magnetic field goes normal, how does the supercurrent stop?}

I am  assuming an ideal situation where the transition is perfectly reversible. For example, for given
applied magnetic field $H<H_c$, the temperature is raised from slightly below $T_c(H)$ to slightly 
above $T_c(H)$. Alternatively, with the system at temperature $T_c(H)$, the magnetic field is
raised from $H-\delta H$ to $H+\delta H$. When the system goes normal, the magnetic field penetrates the body
and the supercurrent stops. Upon slightly cooling or slightly reducing the magnetic field, the 
supercurrent starts flowing again and the magnetic field is expelled.

How the supercurrent stops is 
a highly non-trivial question. In particular, what happens to its mechanical momentum \cite{missing}?
The obvious and only possible answer   is that the momentum of the supercurrent
gets transferred to the body as a whole. For example, if the body is a cylinder hanging from a thread 
with a magnetic field along the direction of its axis, when the supercurrent stops
 the body will start to rotate around its axis.

This experiment has never been performed this way. However an equivalent experiment has been performed.
If a magnetic field is applied to the superconductor, the current starts flowing in one direction and the
body starts rotating in  the same direction. The angular momentum
of the body reflects the angular momentum of the supercurrent, carried by negative electrons moving in direction
opposite to the current flow. In this situation, if the temperature is then raised, both the supercurrent and the
rotation of the body will stop.

But  $how$ is the momentum of the supercurrent transferred to the body as a whole?

The problem is, the process is reversible. The reverse process is the Meissner effect.
Any momentum transfer process involving collisions of electrons in the
supercurrent with phonons or impurities is an irreversible process, hence not allowed.
We argue that the conventional theory of superconductivity $cannot$ answer this question. The question has certainly
never been posed nor answered in the superconductivity  literature.

We have recently posed the question and proposed an answer to it \cite{revers}. The key element of the answer is
$holes$. We argue that {\it the only way} that electrons can transfer mechanical momentum to the body
as a whole in a reversible way is through the motion of $holes$.

\section{Why holes are not like electrons}
The velocity of Bloch electrons is given by
\beq
\vec{v}_k=\frac{1}{\hbar}\frac{\partial \epsilon_k}{\partial \vec{k}}
\eeq
and the acceleration by
\beq
\frac{d \vec{v}_k}{dt}=\frac{1}{\hbar^2}\frac{\partial^2 \epsilon_k}{\partial \vec{k}\partial \vec{k}}\frac{\partial}{\partial t}(\hbar \vec{k})
=\frac{1}{m^*_k}\frac{\partial}{\partial t}(\hbar \vec{k}) .
\eeq
The last equality is for the particular case of an isotropic band, with 
\beq
\frac{1}{m^*_k}\equiv\frac{1}{\hbar^2}\frac{\partial^2\epsilon_k}{\partial k^2} .
\eeq
According to semiclassical transport theory, in the presence of an external force $\vec{F}_{ext}^k$
\beq
\frac{\partial}{\partial t}(\hbar \vec{k}) = \vec{F}_{ext}^k .
\eeq
The total force exerted on a Bloch electron is
\beq
m_e\frac{d \vec{v}_k}{dt}\equiv\vec{F}_{tot}^k=\frac{m_e}{m^*_k} \vec{F}_{ext}^k =  \vec{F}_{ext}^k +
\vec{F}_{latt}^k
\eeq
with $m_e$ the bare electron mass, and $\vec{F}_{latt}^k$ the {\it force exerted by the lattice on the electron} of wavevector $k$, given by
\beq
\vec{F}_{latt}^k=(\frac{m_e}{m^*_k}-1) \vec{F}_{ext}^k
\eeq

Near the bottom of the band $m^*_k$ is positive and $\vec{F}_{latt}^k$ is small. Near the top of the band,
$m^*_k$ is negative and $\vec{F}_{latt}^k$ is larger than  $\vec{F}_{ext}^k$ and points in opposite direction, 
causing the electron near the top of the band to accelerate in direction $opposite$ to the external force.

The importance of this for superconductivity is that when the lattice exerts a force on the electron, 
by Newton's third law the electron exerts a force on the lattice, or in other words transfers momentum
to the lattice. This indicates that the electrons that are most effective in transfering momentum 
from the electrons to the body are electrons near the top of the band. In other words, holes.
This will answer the central question of how the momentum of the supercurrent is transferred
to the body as a whole in a reversible way, without  energy dissipation.

\section{how holes answer the central question in superconductivity}

    \begin{figure} [htb]
 \resizebox{8.5cm}{!}{\includegraphics[width=6cm]{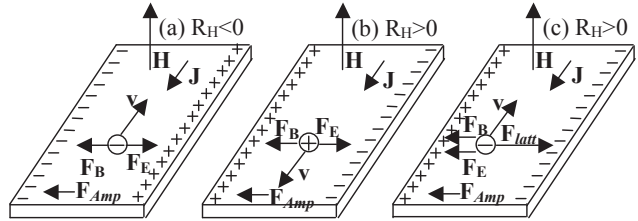}}
 \caption { Hall effect for a material with negative Hall coefficient (a) and for a material with positive
 Hall coefficient (b), (c).  $F_B$ and $F_E$ are the magnetic and electric Lorentz forces acting on
 carriers, $F_{latt}$ is the force exerted by the lattice on the electron,  $\vec{J}$ is the current density,
 $\vec{F}_{Amp}$ is the Amperian force on the bar.}
 \label{figure1}
 \end{figure} 
 
 Figure 8 shows the balance of forces on carriers   in Hall bars with negative and positive Hall coefficients.
 For $R_H<0$,  (Fig. 8 (a)), electric and magnetic forces on electrons are balanced, and 
 for $R_H>0$,  (Fig. 8 (b)), electric and magnetic forces on holes are balanced.
 However, in that case, electric and magnetic forces on electrons are $not$ balanced, as shown in 
 Fig. 8 (c)). For electrons to propagate along the direction of the current $\vec{J}$, another force
 is needed to balance electric and magnetic forces when $R_H>0$. That is the force exerted by the lattice on
 electrons, $F_{latt}$.
 
In the following, the forces under discussion are understood to be in direction perpendicular to the flow of
the current $\vec{J}$  in Fig. 8. (There is also an electric force in direction parallel to $\vec{J}$  that is of no interest
for the issue at hand). It is easy to see that the total force exerted by the lattice on the carriers is zero
 for a band close to empty with $R_H<0$ and is not zero for 
 a band close to full and $R_H>0$: the total force exerted by both the
 lattice and the external fields on the current carrying carriers has to be zero, hence from  Eq. (5)
 \beq
\sum_{occ}\vec{F}_{tot}^k=\sum_{occ} \frac{m_e}{m^*_k} \vec{F}_{ext} ^k=0
\eeq
where the sum is over occupied $k$ states. 
 For the case $R_H<0$ and the band  close to empty we can assume that the effective mass
 is independent of $k$, $m^*_k=m^*$. From Eq. (7)
 \beq
 \sum_{occ} \frac{m_e}{m^*_k} \vec{F}_{ext}^k=
 \frac{m_e}{m^*} \sum_{occ} \vec{F}_{ext}^k=0
 \eeq
 therefore 
 \beq
  \sum_{occ} \vec{F}_{ext}^k=0
  \eeq
  and Eqs. (6), (8) and (9) imply
  \beq
 \sum_{occ}\vec{F}_{latt}^k=0
 \eeq
 so that the total force exerted by the lattice on the carriers in direction perpendicular to the 
 current flow is zero, and so is the total force exerted by the carriers on the lattice.
 
 Instead, for a band that is close to full and $R_H>0$, we cannot assume that 
 $m^*_k$ is independent of $k$ for the occupied states, instead we assume
 $m^*_k=-m^*$ for the empty states. Eq. (7) then implies       
 \beq
\sum_{occ}\vec{F}_{tot}^k=-\sum_{unocc}\frac{m_e}{m^*_k} \vec{F}_{ext} ^k= - \frac{m_e}{m^*}
\sum_{unocc} \vec{F}_{ext} ^k=0
\eeq
and from Eqs. (6) and (11)
 \beqn
  \sum_{occ}\vec{F}_{latt}^k&=&-\sum_{occ}\vec{F}_{ext}^k
  \\ \nonumber
  &=&
  -\sum_{all}\vec{F}_{ext}^k+\sum_{unocc}\vec{F}_{ext}^k=
  -2N e \vec{E} \neq 0  
  \eeqn
  where $N$ is the number of $k-$points in the Brillouin zone. To obtain Eq. (12) we used that
  the external force is
  \beq
  \vec{F}_{ext}^k=e\vec{E}+\frac{e}{c}\vec{v}_k\times\vec{B}
  \eeq
  with $\vec{E}, \vec{B}$ the electric and magnetic fields.
  
  Eq. (12) shows that when $R_H>0$ the lattice exerts a force on the conducting carriers that is
  perpendicular to the current flow.
  Conversely, the conducting carriers exert a force on the lattice or, in other words,
  transfer momentum to the lattice in direction perpendicular to the current flow. 
  This force on the lattice, plus the electrostatic force on the positive ions that
  points in opposite direction (to the right in Fig. 8 (c)) gives the Amperian force on the
  Hall bar, $\vec{F}_{Amp}$.
  In contrast, if the carriers are electrons with $R_H<0$,
  the same $\vec{F}_{Amp}$ results from the direct force of the electric field on the ions
  (to the left in fig. 8 (a)) and there is no net force exerted by electrons on the lattice nor by 
  the lattice on electrons 
  in direction perpendicular to the current flow, hence no momentum transfer from the carriers to the lattice.

 This is, in essence, why hole carriers are indispensable for superconductivity  \cite{momentum}.
  Let us next discuss how this explains the process of momentum transfer from the
 supercurrent to the body when the supercurrent stops.
 
      \begin{figure} [htb]
 \resizebox{8.5cm}{!}{\includegraphics[width=6cm]{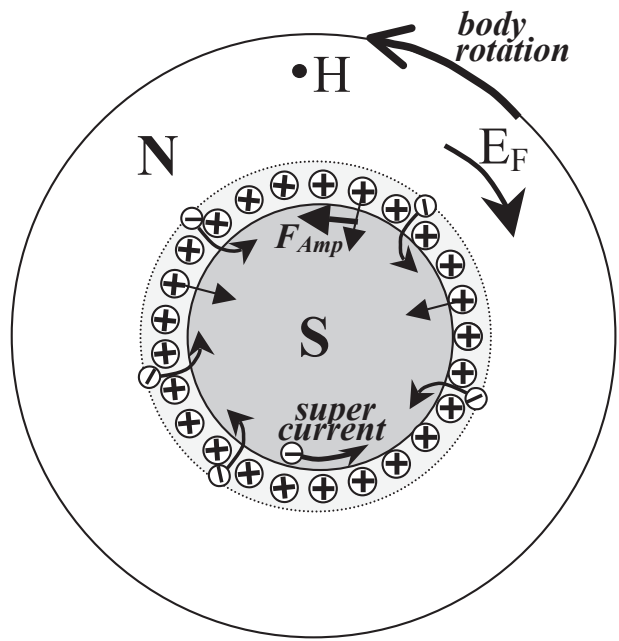}}
 \caption { Superconductor to normal transition in a magnetic field $H$  pointing out of the paper. 
 Supercurrent flows clockwise, electrons carrying supercurrent move counterclockwise. The inward motion of the
 phase boundary is accompanied by inward flow of negative charge, that stops the 
 supercurrent flow, and inward flow of normal hole carriers, that transfers the
 angular momentum of the supercurrent to the body as a whole that acquires counterclockwise rotation.
 Details are given in ref. \cite{disapp}.}
 \label{figure1}
 \end{figure} 
 
 Figure 9 shows schematically how the supercurrent stops when a cylindrical superconductor in a magnetic
 field pointing out of the paper goes normal. The inward motion of the N/S phase boundary is accompanied by a radial flow
 and counterflow of charge. In the process of becoming normal, superconducting electrons flow
 inward and are stopped by the clockwise magnetic Lorentz force resulting from this radial motion.
 At the same time normal holes flow inward and exert a torque on the body in the counterclockwise direction.
 The force $\vec{F}_{Amp}$ shown in the figure pointing counterclockwise is
 the same as the Amperian force in figure 7 (c) pointing left.
 In that way, the counterclockwise angular momentum possessed by the supercurrent is transferred to
 the body as a whole without involving irreversible collisions that would dissipate Joule heat.
 In the reverse process where a normal cylinder becomes superconducting and expels the magnetic 
 field, the direction of the motions in Fig. 9 are simply reversed (except for the direction of the supercurrent).
 The details of these processes are discussed in references \cite{disapp,momentum}.

 \section{discussion}

Alex M\"uller's and George Bednorz's breakthrough discovery in 1986 ushered in the era of hole superconductivity.
Before their discovery, the term `hole superconductor' had never been used, after their discovery
it became commonplace. The evidence that hole carriers are necessary for superconductivity
continues to accumulate. Of course sometimes it is the case that in a multiband situation electron
carriers exist and dominate the transport, in which case it may not be obvious that hole carriers
also exist and are responsible for superconductivity.

Possibly the one example where it is least obvious that hole carriers exist is for the
very low carrier density n-doped semiconductor  $SrTiO_3$ \cite{sr1,sr2,srti}.
It is believed that only electron carriers exist in this material, however according to the physics discussed in this paper superconductivity is impossible without hole carriers. Therefore, we conjecture that there is
at least two-band conduction in this material \cite{sr2} with one of the bands hole-like, the hole carriers would
be induced by electron doping just like in the case of the electron-doped cuprates \cite{holesedoped}.
It is interesting that Alex M\"uller and coworkers in 1976 reported the finding of trapped holes  near dopant impurities  \cite{srtialex},
which suggests that holes are easily induced  in this material.

30 years after the discovery of the cuprate superconductors it is becoming increasingly  clear 
that {\it all superconductors are hole superconductors}  \cite{1991}. There are many reasons for
this, all interconnected \cite{holesc}. I believe the most fundamental  reason is
the one discussed in this paper, which can be summarized in the following very simple statements:
\begin{itemize}
\item In any superconductor, the mechanical momentum of the supercurrent  has to be transferred to the body as a whole when the supercurrent stops.
\item The process is reversible under ideal conditions.
\item Only {\it hole carriers} can transfer mechanical momentum from electrons to the body as a whole in a reversible way, electron carriers cannot.
\end{itemize}

If this is so, it had been glimpsed at in the early days of superconductivity \cite{chapnik} but then was 
 well hidden from superconductivity researchers for a long time,
buried under the heavy weight of BCS theory, until
Alex M\"uller's and George Bednorz's 1986   discovery of high $T_c$ superconductivity in 
cuprates started the process of
bringing this deep secret of superconductors again into the open. 

I am extremely grateful to them for having led me to this understanding. Happy 90th birthday Alex!

\end{document}